# A dynamic plasmonic system that responds to thermal and aptamer-target regulations

Chao Zhou[1,†], Ling Xin[1], Xiaoyang Duan[1,2], Maximilian J Urban[1,2], and Na Liu[1,2*]

[1]Max Planck Institute for Intelligent Systems, Heisenbergstrasse 3, D-70569 Stuttgart, Germany
[2]Kirchhoff Institute for Physics, University of Heidelberg, Im Neuenheimer Feld 227, D-69120, Heidelberg, Germany

The DNA origami technique has empowered a new paradigm in plasmonics for manipulating light and matter at the nanoscale. This interdisciplinary field has witnessed vigorous growth, outlining a viable route to construct advanced plasmonic architectures with tailored optical properties. However, so far plasmonic systems templated by DNA origami have been restricted to respond to only single stimuli. Despite broad interest and scientific importance, thermal and aptamer-target regulations have not yet been widely utilized to reconfigure three-dimensional plasmonic architectures. In this Letter, we demonstrate a chiral plasmonic nanosystem integrated with split aptamers, which can respond to both thermal and aptamer-target regulations. We show that our dual-responsive system can be noninvasively tuned in a wide range of temperatures, readily correlating thermal control with optical signal changes. Meanwhile, our system can detect specific targets including adenosine triphosphate and cocaine molecules, which further enhance the optical response modulations and in turn influence the thermal tunability.

Keywords: Plasmonics, DNA origami, Thermal control, Aptamer-target interactions, Circular dichroism, Chirality

One of the most powerful workhorses in DNA nanotechnology is the so-called DNA origami technique, which was invented by Paul Rothemund in 2006.[1] DNA origami can be formed by folding a long scaffold DNA strand with hundreds of short staple strands into nearly any arbitrary shapes.[1-3] Due to the unique sequence specificity, DNA origami is fully programmable and addressable, offering unprecedented precision for organization of metal nanoparticles in space.[4-6] With this approach, a wealth of plasmonic architectures with interesting optical phenomena has been demonstrated,[7,8] including plasmon hybridization,[9] magnetic resonances,[10] chiral responses,[11-19] as well as surface-enhanced fluorescence[20,21] and Raman scattering[22-25].

Along with the fast development of this interdisciplinary field, the inevitable evolution from static to dynamic plasmonic architectures has been envisaged.[26] The DNA origami technique poses a great opportunity to endow plasmonic systems with dynamic functionalities. Different schemes for structural reconfigurations can be employed, for instance, toehold-mediated strand displacement reactions[27-31], RNA[32], pH[33,34], and light stimuli[33,35]. As a result, dynamic plasmonic systems, which can perform diverse mechanical motions including rotation[36], walking[28,29], and sliding[37] have been successfully implemented. This significantly advances DNA nanotechnology to a new dimension of functional potential. However, there is still plenty of room to enrich the toolbox of DNA-based dynamic plasmonics and nanomachinery by exploring novel external inputs, for instance, thermal and aptamer-target regulations. Thermal stimuli are a noninvasive means to impart reconfigurability[38,39] to plasmonic systems without producing any chemical waste. Aptamers are artificial DNA or RNA oligonucleotides selected from random-sequence nucleic acid libraries. They can bind to various specific targets with high selectivity and affinity, extremely useful in clinic diagnostics.[40] Therefore, aptamer-functionalized plasmonic systems hold a great potential to mediate target-responsive optical signal transduction, suggesting a new generation of optical biosensors for detection of various target molecules.

Figure 1a shows the schematic of our dual-responsive plasmonic nanosystem. The DNA origami template comprises a rotary bundle, which connects to a two-layer rectangular plate at its center via the scaffold DNA strand. The details of the origami design can be found in Supporting Information Fig. S1 and Table S1. One gold nanorod (AuNR) is assembled on the rotary bundle and the other one is immobilized on the back surface of the origami plate. Both of the AuNRs have dimensions of 38 nm × 10 nm. The crossed AuNRs constitute a 3D plasmonic chiral object. When interacting with circularly polarized light, the excited plasmons in the two AuNRs can be coupled owing to their close proximity, giving rise to strong circular dichroism (CD) spectra. Here a chiral platform is adopted in that it allows for optical differentiation of sequential configuration changes executed by a dynamic plasmonic system in real time.[26]

Two types of split aptamers, *i.e.*, split adenosine triphosphate (ATP)[41,42] and cocaine (COC) aptamers[43,44] are arranged on the DNA origami template, either separately or jointly. Each split aptamer consists of two DNA strands. One is extended from the rotary bundle and the other one is from the origami plate as shown in Fig. 1a. More specifically, a pair of split ATP aptamers is positioned along the *a*-*a* diagonal direction (see the black-dashed line). The two strands of each split ATP aptamer are located at *a* on the origami plate and $\bar{a}$ on the rotary bundle, respectively (also see Fig. 2a). If the rotary bundle is bound to the origami plate along the *a*-*a* diagonal direction, the two AuNRs form a left-handed (LH) state. On the other side, a pair of split COC aptamers is positioned along the *b*-*b* diagonal direction (see blue-dashed line in Fig. 1a). One strand is located at *b* and the other one $\bar{b}$ is extended from the rotary bundle in each split COC aptamer (also see Fig. 3a). If the rotary bundle is bound to the origami plate along the *b*-*b* diagonal direction, the two AuNRs form a right-handed (RH) state. Representative transmission electron microscopy (TEM) images of the DNA origami templates and the plasmonic nanostructures are presented in Figs. 1b and 1c, respectively, demonstrating a high fidelity of the assembly. The overview TEM images of the structures are shown in Supporting Information Fig. S2. It is noteworthy that the concentration of the plasmonic nanostructures in each system is estimated using the AuNR absorbance at their longitudinal resonance (~762 nm). To ensure quantitative comparisons, the concentration of the plasmonic system in each experiment is tuned to 0.6 nM.

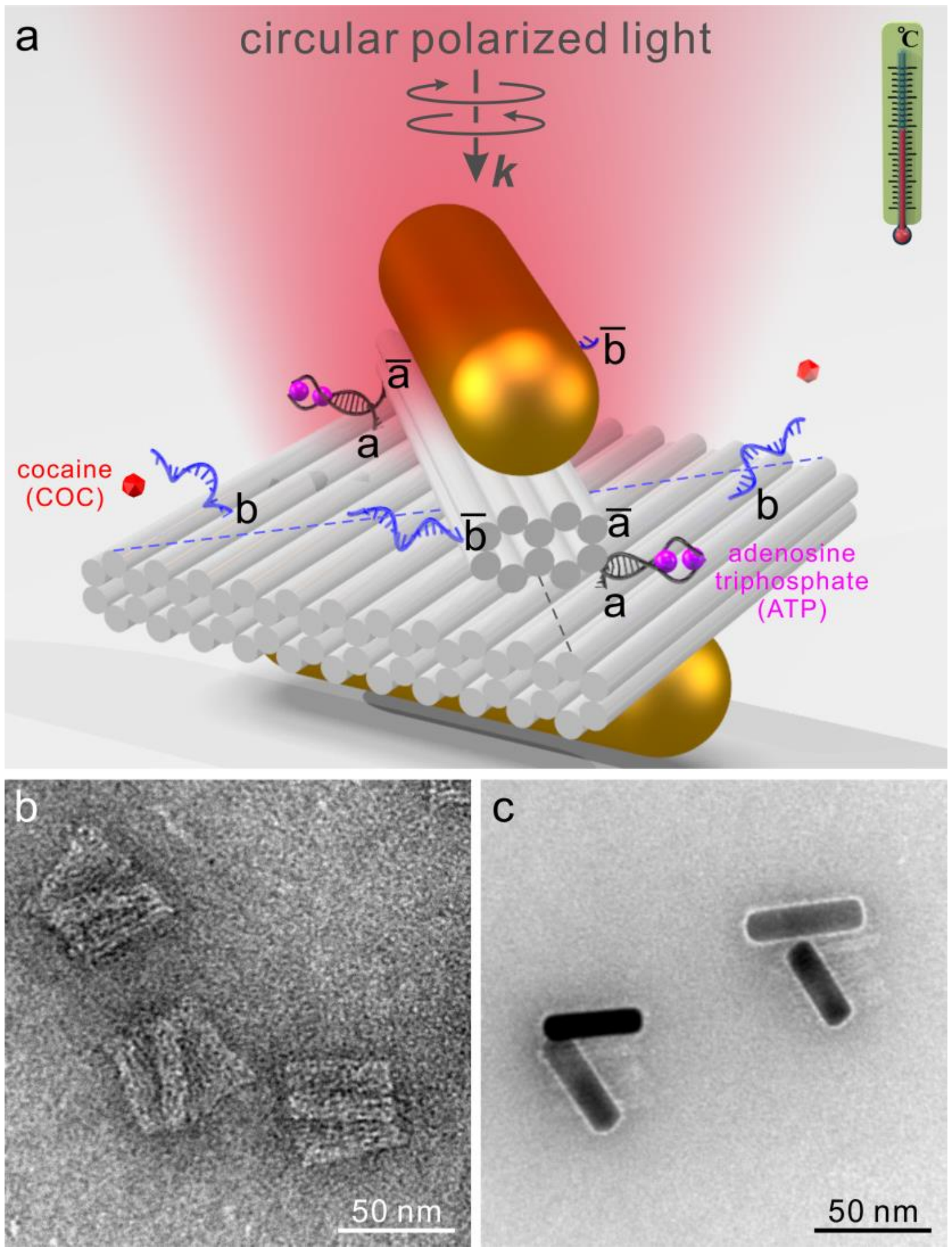

**Figure 1.** (a) Schematic of the dual-responsive plasmonic nanosystem that can respond to thermal and aptamer-target regulations. The magenta spheres represent adenosine triphosphate (ATP) molecules and the red polyhedra represent cocaine (COC) molecules. TEM images of the DNA origami templates (b) and the assembled plasmonic nanostructures (c).

We first demonstrate dynamic regulations of the plasmonic system by the split ATP aptamers as illustrated in Fig. 2a. In this case, the split COC aptamers are absent on the origami. The two separate strands of each split ATP aptamer possess a 5 base-pair (bp) hybridization segment (see the inset of Fig. 2a). To investigate the thermal characteristics, the CD spectra of the sample are measured using a Jasco-1500 CD spectrometer with a Peltier temperature controller in a temperature range of 5 to 35 °C (2 °C/step). As shown in Fig. 2b, the chiroptical responses of the sample sensitively depend on the temperature changes. More specifically, the CD strength becomes more pronounced, when the temperature decreases. The results provide rich insights into understanding of the thermal tunability. A 5 bp hybridization segment in the split ATP aptamer is too short to be stable at room temperature (25 °C). Two of such aptamer motifs arranged on the origami lead to a metastable system, in which only a fraction of the nanostructures can be locked into the LH state. Upon a temperature decrease, the segment becomes more stabilized so that more structures can be directed to the LH state, giving rise to a CD increase. In contrast, upon a temperature increase, the segment is destabilized with more structures in the relaxed state, resulting in a CD decrease. The thermal tuning is carried out using a Peltier temperature controller, which takes on average 1 min to complete a 2 °C temperature change. The samples are measured, once a designated temperature is reached. The switching time is much shorter than that of the DNA-fueled plasmonic system.[27] This is majorly because in the present case locking of the structures is not limited by the diffusion kinetics of the added DNA fuel strands. Rather, the split aptamer segments are positioned in close proximity on the origami, resulting in a high local concentration of the complementary segments and thus facilitating the dynamic responses.

Thermal regulation features a noninvasive means to dynamically reconfigure plasmonic nanostructures. To evaluate the reversibility, cyclic operation of the sample is carried out between 15 °C and 35 °C. The CD intensities at a fixed wavelength of 730 nm are recorded in different cycles and shown in Fig. 2c. There is no clear signal degradation observable, demonstrating a robust switching performance. This indicates that the plasmonic nanostructures retain their structural integrity during cycling. The result also confirms that the observed CD switching stems from the hybridization/dehybridization of the split aptamers upon thermal tuning. The influence of the bp number in the segment on the thermal characteristics is presented in Supporting Information Figs. S3-S5.

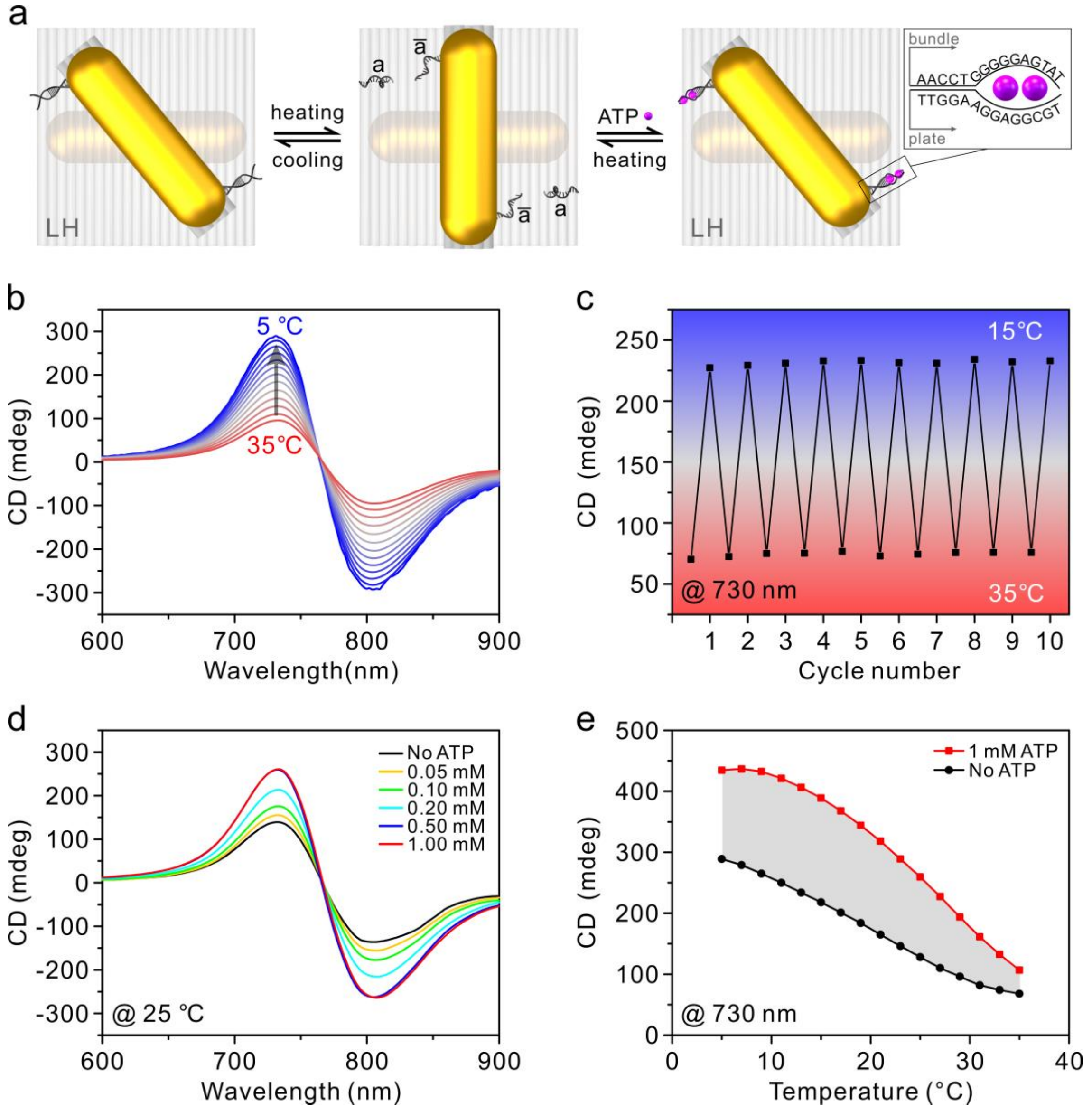

**Figure 2.** (a) Schematic of the thermal and split ATP aptamer-target regulations of the plasmonic system. (b) Temperature-dependent CD spectra (2 °C/step). (c) Cyclic operation of the plasmonic system between 15 ºC and 35 °C. (d) CD responses in dependence on the ATP concentration. (e) Comparison of the thermal characteristics in the absence and presence (1 mM) of the ATP molecules.

Next, we demonstrate dynamic regulations of the plasmonic system by the split ATP aptamers upon target binding. Each split aptamer can bind to two ATP molecules.[45] Before ATP binding, the CD spectrum measured at 25 °C is presented by the black curve in Fig. 2d. Upon addition of 0.05 mM ATP molecules, the CD intensity exhibits an obvious increase (orange curve), suggesting that more plasmonic nanostructures are locked into the LH state. With further ATP concentration increases, the chiroptical responses continue to be enhanced and reach a stable state at an ATP concentration of 0.5 mM. A higher concentration (1 mM) does not give rise to a further CD intensity increase. To investigate how ATP binding will in turn influence the thermal characteristics of the system, the CD intensities at 730 nm are recorded as a function of temperature in the absence and presence (1 mM) of the ATP molecules, respectively (see Fig. 2e). It is apparent that ATP binding results in overall stronger chiroptical responses. At the same corresponding temperatures, more plasmonic nanostructures can be locked, corroborating the designated LH state, when compared to the cases without ATP molecules. Meanwhile, ATP binding also enhances the stability of the chiral plasmonic system in response to thermal tuning (see Supporting Information Figs. S6-S8).

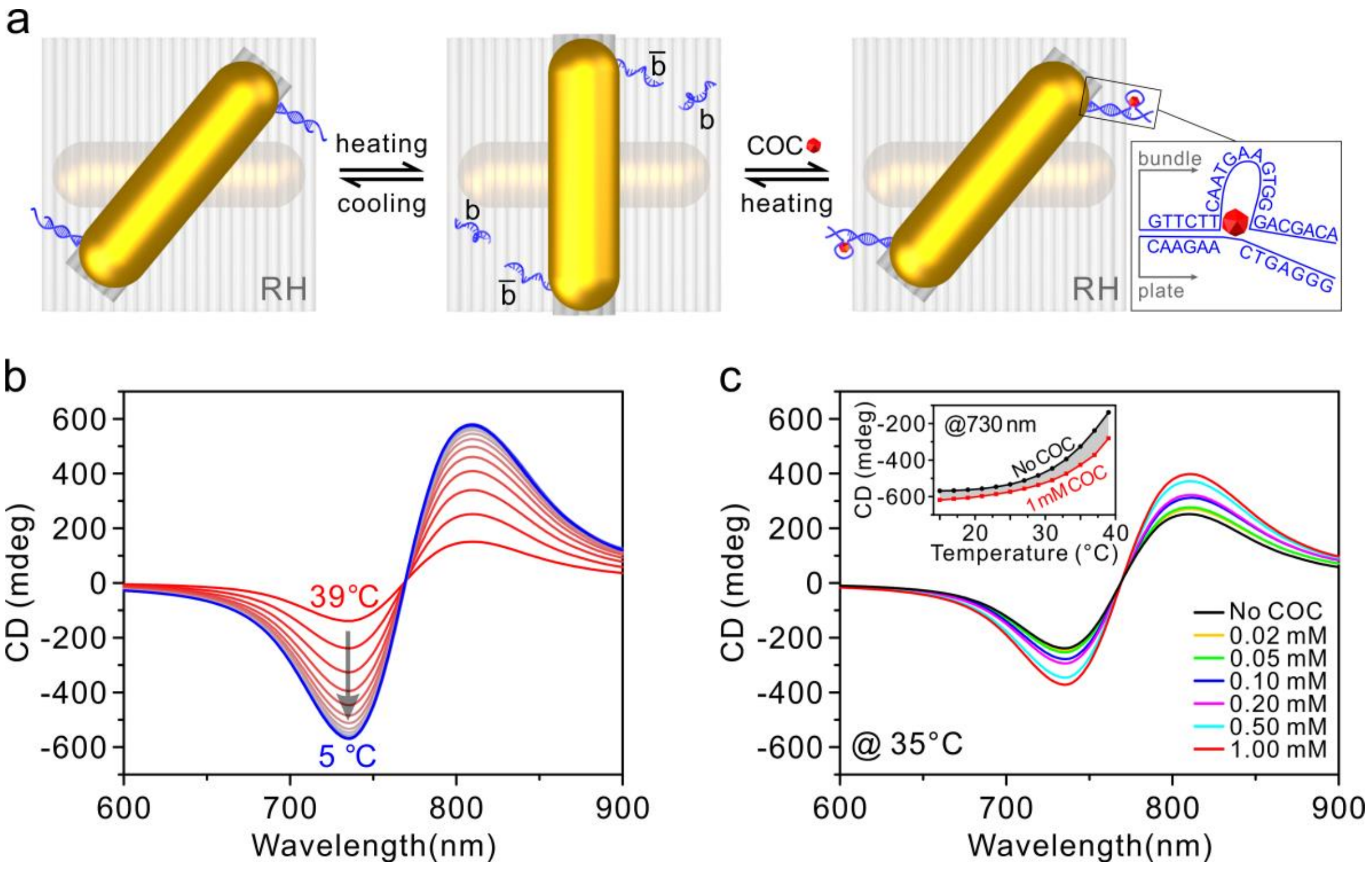

**Figure 3.** (a) Schematic of the thermal and split COC aptamer-target regulations of the plasmonic system. (b) Temperature-dependent CD spectra (2 °C/step). (c) CD responses in dependence on the COC concentration. Inset: Comparison of the thermal characteristics in the absence and presence (1 mM) of the COC molecules.

Taking a further step, dynamic regulations of the plasmonic system by the split COC aptamers are investigated (see Fig. 3a). In this case, the split ATP aptamers are absent on the origami. Each split COC aptamer consists of two separate strands with a 6 bp hybridization segment. As they are arranged along the *b*-*b* diagonal (opposite to the *a*-*a* direction), the locked plasmonic nanostructures are in the RH state. The generated CD spectra are thus mirrored, when compared to those obtained from the split ATP aptamers. Similar to the case in Fig. 2b, the chiroptical responses can be gradually enlarged by decreasing temperature as shown in Fig. 3b. Furthermore, dynamic regulations upon COC binding exhibit clear concentration dependence. Higher COC concentrations lead to stronger chiroptical responses as shown in Fig. 3c. In addition, COC binding enhances the stability of the chiral plasmonic system in response to thermal tuning (see the inset of Fig. 3c). The thermal characteristics in dependence on the split COC aptamer number arranged on the origami template can be found in Supporting Information Fig. S9.

Finally, we examine dynamic regulations of the plasmonic system in the presence of both split ATP and COC aptamers upon target binding as illustrated in Fig. 4a. To balance the binding affinities of these two different aptamers at 35 °C through evaluating the respective CD values, split ATP aptamers with 6-bp hybridization segments are utilized (see Supporting Information Figs. S4 and S7), while the split COC aptamers remain unchanged (see Fig. 3). As shown by the black curve in Fig. 4b, before target binding the plasmonic system exhibits a RH preference. This agrees with the previous observation that the split COC aptamer-locked RH state (see Fig. 3b) yields a larger absolute CD value than that from the split ATP aptamer-locked LH state at the same concentration and temperature (see Fig. S4). Upon addition of the ATP molecules (1 mM), the chiroptical responses are flipped (red curve). This elucidates that ATP binding helps to direct plasmonic nanostructures to the LH state. On the contrary, when COC molecules are added, the plasmonic nanostructures are driven more effectively to the RH state, showing stronger RH chiroptical responses (blue curve). To this end, our plasmonic system can sense multiple target species and translate the specific binding events into optical signal changes.

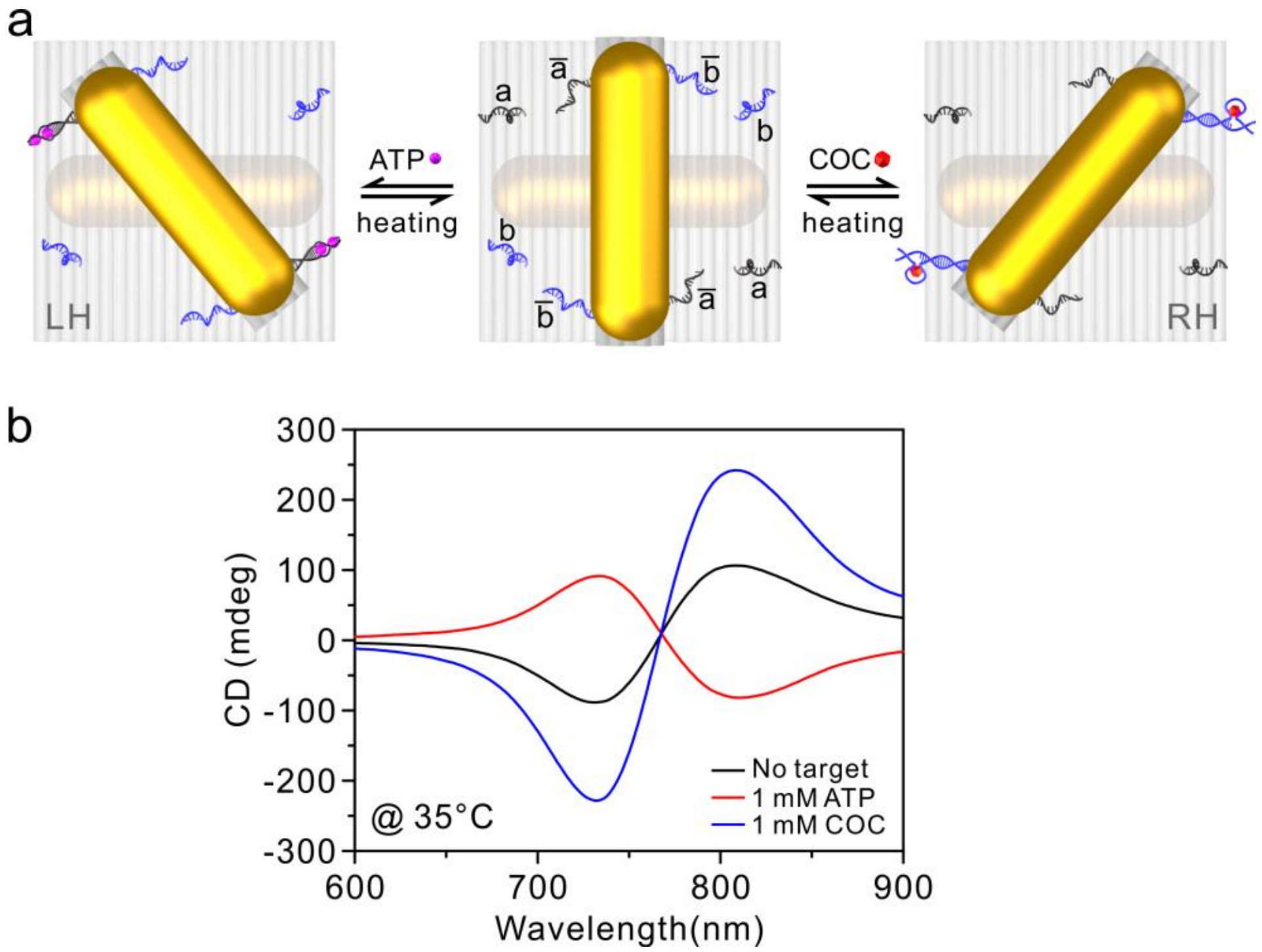

**Figure 4.** (a) Dynamic regulations of the plasmonic system in the presence of both split ATP and COC aptamers upon target binding. (b) CD spectra without target binding and upon additions of ATP (1 mM) and COC (1 mM) molecules, respectively.

In conclusion, we have demonstrated a dual-responsive plasmonic system, which can be regulated by both thermal control and aptamer-target interactions. Aptamer-functionalized plasmonic systems templated by DNA origami can undergo concomitant structural reconfigurations and therefore report optical response changes upon thermal tuning and target binding. In turn, such systems also feature a novel plasmonic sensing platform, taking the unprecedented programmability and addressability afforded by DNA origami as well as the high specificity and selectivity of aptamer-target interactions. Our concept can be readily adopted to develop more universal dynamic plasmonic architectures, which can respond to various selective molecular binding events. This will hold a great potential to detect different molecular targets with high affinity and specificity for clinic diagnostics and bioanalytical applications.

ASSOCIATED CONTENT

**Supporting Information**
Experimental details, TEM images, chiroptical analysis, DNA origami design and sequences.

AUTHOR INFORMATION

**Corresponding Author**
*na.liu@kip.uni-heidelberg.de

**Notes**
The authors declare no competing financial interest.

**Author Contributions**
C.Z. and N.L. conceived the project. C.Z. and L.X. performed the experiments. C.Z. and N.L. wrote the manuscript. X.D. contributed to the scientific visualization. All authors discussed the results and analyzed the data. All authors have given approvals to the final version of the manuscript.

**Present Address**
†CAS Key Laboratory of Nano-Bio Interfaces, Division of Nanobiomedicine and *i*-Lab, Suzhou Institute of Nano-Tech and Nano-Bionics, Chinese Academy of Sciences, Suzhou 215123, P. R. China.

ACKNOWLEDGMENT

This project was supported by the Sofja Kovalevskaja grant from the Alexander von Humboldt-Foundation, the Volkswagen foundation, and the European Research Council (ERC Dynamic Nano) grant. M.U. acknowledges the financial support by the Carl-Zeiss-Stiftung. We thank Marion Kelsch for assistance with TEM. TEM images were collected at the Stuttgart Center for Electron Microscopy.

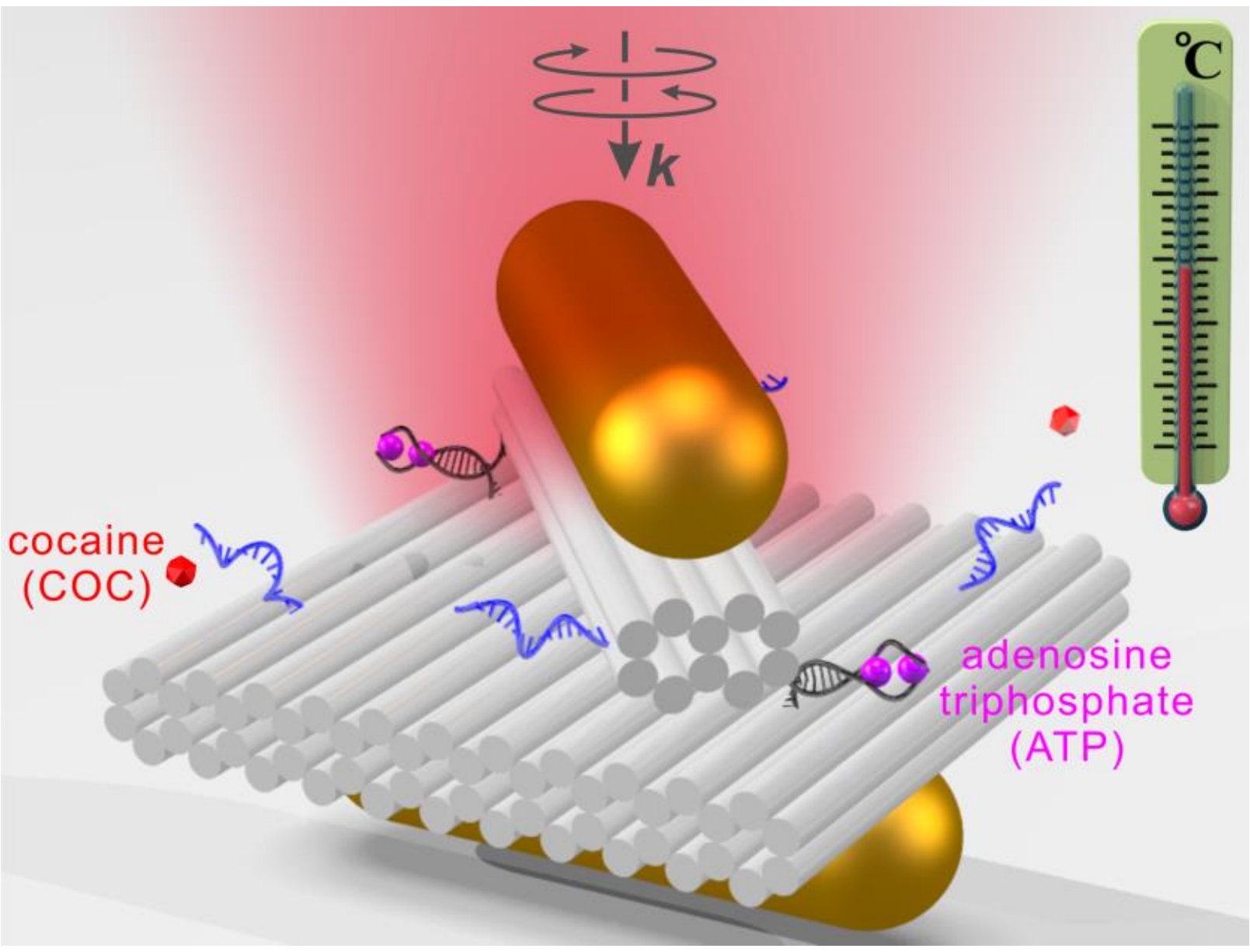

TOC